\documentclass[twocolumn,prl,aps]{revtex4}
\usepackage{amssymb}
\usepackage{epsfig}

\begin{document}
\title{Fractal geometry of spin-glass models}
\author{J.\ F.\ Fontanari}
\affiliation{Instituto de F{\'\i}sica de S{\~a}o Carlos,
  Universidade de S{\~a}o Paulo,
  Caixa Postal 369, 13560-970 S\~ao Carlos SP, Brazil}
\author{P.\ F.\ Stadler}
\affiliation{Institut f{\"u}r Theoretische Chemie,
       Universit{\"a}t Wien,
       W{\"a}hringerstra{\ss}e 17, A-1090 Wien, Austria\\
       The Santa Fe Institute,
       1399 Hyde Park Road, Santa Fe, NM 87501, USA}
\begin{abstract}
Stability and diversity are two key properties that living entities share
with spin glasses, where they are manifested through the breaking of the
phase space into many valleys or local minima connected by saddle
points. The topology of the phase space can be conveniently condensed into
a tree structure, akin to the biological phylogenetic trees, whose tips are
the local minima and internal nodes are the lowest-energy saddles
connecting those minima.  For the infinite-range Ising spin glass with
$p$-spin interactions, we show that the average size-frequency distribution
of saddles obeys a power law $\langle \psi(w) \rangle \sim w^{-D}$, where
$w=w(s)$ is the number of minima that can be connected through saddle $s$,
and $D$ is the fractal dimension of the phase space.
\end{abstract}
%\pacs{75.10.Nr, 87.23.Kg}
\maketitle
\phantom{.}
\vspace{-1.0 cm}
The resemblance between models of adaptive evolution \cite{Kauffman_1} and
disordered spin systems \cite{MPV} is certainly not coincidental. In fact,
two key features that any successful model of biological evolution ought to
possess are stability and diversity, i.e., exactly the properties
responsible for the complex thermodynamics of spin glasses
\cite{Anderson}. It is not a surprise therefore, that many of the tools
and concepts of the statistical mechanics of disordered systems have been
applied to the study of the evolutionary process. In this contribution we
show that such interchange can be profitable to statistical mechanics too,
in that a great deal of information about the phase space of spin-glass
models can be condensed into a tree structure, in a standard procedure
widely used in taxonomy and molecular phylogenetics \cite{TreeBooks}. As in
the biological case, the geometric properties of this tree can be used to
characterize the disordered system quantitatively.

%%% Landscape Models
The main unifying concept in the investigations of the physics of disordered
systems and evolutionary change is probably the
notion of fitness or energy landscape.
The concept of neighborhood among genotypes (configurations),
typically defined such that point mutations interconvert neighbors, allows
us to view the set of all genotypes as the vertices of a graph with edges
connecting neighboring configurations. A {\em fitness landscape} is then
obtained by assigning a fitness value to each vertex.  An explicit connection
between those two research fields is obtained in the case
of populations of asexually reproducing haploid organisms evolving on
rugged fitness landscapes.  In this case the genotypes are often
modeled by configurations of $N$ Ising spins $s=(s_1,\ldots,s_N)$ with
$s_i=\pm1$ so that a point mutation corresponds to a single spin flip.
In the simplest
case, evolutionary adaptation is described as an ``adaptive walk''
on the fitness landscape \cite{Kauffman_1}, whose statistical mechanics
equivalent is the zero-temperature Glauber dynamics. 
The fitness function assigns a random numerical value to each one of the $2^N$
spin configurations. In this work we consider the $p$-spin landscapes
\cite{Derrida}:
\begin{equation}\label{H_p}
 \mathcal{H}_p(s) =
- \sum_{1 \leq i_1 \leq  i_2 \ldots \leq i_N \leq N}
 J_{i_1 i_2 \ldots i_N } s_{i_1} s_{i_2} \ldots s_{i_N}
\end{equation}
where the $J_{i_1 i_2 \ldots i_N }$ are statistically independent Gaussian
distributed random variables with mean zero and variance $p\,!/(2
N^{p-1})$. 
The $p$-spin models form a class of tunably rugged landscapes
similar to Kauffman's Nk-model \cite{Kauffman_1}, which is not only more
appealing to statistical mechanics but also is a more natural basis of
landscape theory \cite{Stadler:96b}. In fact, for $p=2$ the Hamiltonian
${\mathcal{H}}_p$ reduces to the SK model \cite{Sherrington} which exhibits
a large number of highly correlated local minima, while the limit
$p\to\infty$ corresponds to the random energy model (REM) \cite{Derrida}
and yields an extremely rugged, uncorrelated landscape.  
Like the Nk-model,
$p$-spin landscapes have been used repeatedly to model evolutionary
processes, see e.g.\ \cite{Anderson,evoPspin}.

%%% Basins and Barriers 
The scenario that emerges from the replica approach to disordered spin
systems is that the phase space composed of the $2^N$ spin configurations
is broken into many valleys \cite{MPV}. The ease with which one valley can
be reached from another one depends on the saddle points connecting
them. More specifically, the energy of the lowest saddle point separating
two local minima  $x$ and $y$ is
\begin{equation}
  E[x,y] = \min_{ \mathbf{p}\in\mathbb{P}_{xy} } \,
           \max_{ z \in \mathbf{p} } {\mathcal{H}}_p (z)
\label{eq:saddle}
\end{equation}
where $\mathbb{P}_{xy}$ is the set of all paths $\mathbf{p}$
connecting $x$ and $y$ by a series of subsequent spin-flips (or
point mutations).  The saddle-point energy $E[\,.\,,\,.\,]$ is an
ultrametric distance measure on the set of local minima, see e.g.\ 
\cite{ultrametric}.

\begin{figure}[t]
\centerline{\epsfig{width=0.49\textwidth,file=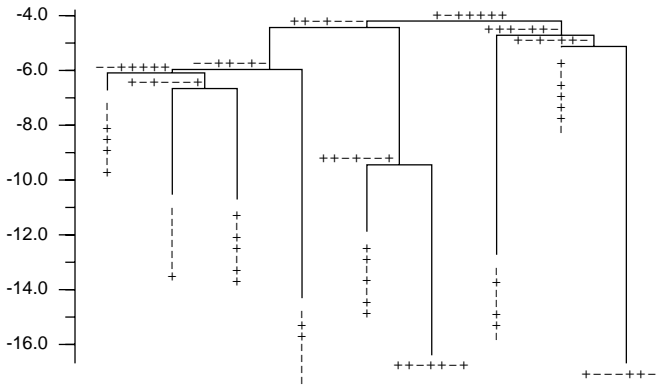}}
\par
\caption{Barrier tree for a 3-spin model with $N=7$. The 9 minima (tips)
and 8 connecting saddle points (internal nodes) are labeled by their spin
configurations.}
\label{fig:3spin}
\vspace{-0.4truecm}
\end{figure}

Let us assume for a moment that the energy function is non-degenerate,
i.e., ${\mathcal{H}}_p (x)\ne {\mathcal{H}}_p (y)$ whenever $x\ne y$. 
This is true for generic $p$-spin
models with odd interaction order $p$. Then there is a unique saddle point
$s=s(x,y)$ connecting $x$ and $y$ characterized by
${\mathcal{H}}_p(s)=E[x,y]$. Note that this definition of saddle point is more
restrictive than in differential geometry where saddles are not required to
separate local optima \cite{Vertechi:89}. To each saddle point $s$ there is
a unique collection of configurations $B(s)$ that can be reached from $s$
by a path along which the energy never exceeds $\mathcal{H}_p(s)$. In other
words, the configurations in $B(s)$ are mutually connected by paths that
never go higher than $\mathcal{H}_p(s)$. This property warrants to call
$B(s)$ the {\em valley or basin below the saddle $s$}. Furthermore,  suppose
that $\mathcal{H}_p(s)<\mathcal{H}_p(s')$. Then there are two possibilities:
if $s\in B(s')$ then $B(s)\subseteq
B(s')$, i.e., the basin of $s$ is a ``sub-basin'' of $B(s')$, or $s\notin
B(s')$ in which case $B(s)\cap B(s')=\emptyset$, i.e., the valleys are
disjoint. This property arranges the local minima and the saddle points in
a unique hierarchical structure which is conveniently represented as a
tree, termed {\it barrier tree} (see Fig.~\ref{fig:3spin}).

In principle, barrier trees can be computed by means of the following
simple recursive procedure: The tips of the tree are the local minima.  The
parent of tip $x$ is the lowest-energy saddle point $s$ that connects $x$
to another local minimum.  Analogously, the parent of a saddle point $s$ is
another saddle point $s'$ that connects $s$ to a local minimum $z$ that is
not contained in the basin below $s$, i.e., $z\notin B(s)$.  The ``root''
of the resulting tree is the saddle point $s^*$ with the highest energy,
since by definition all local minima are contained in $B(s^*)$. Note that
the subtree $\mathfrak{T}(s)$ that has the saddle $s$ as its root has
exactly the local minima in $B(s)$ as its tips.  The exact calculation of
the barrier tree is a highly challenging computational problem and only
recently some progress in that direction has been achieved, mainly in the
context of RNA and protein folding \cite{Flamm,Garstecki} (see also
\cite{prevBT}). The reason is that, unless one has sufficient {\it a
priori} knowledge on the landscape, it is necessary to generate the
complete landscape in order to find all local minima.  Even for very small
system sizes a simple-minded exhaustive search approach to evaluating
Eq.~(\ref{eq:saddle}) would be hopeless as one must calculate all paths
connecting all pairs of minima.  In this contribution we use the program
package {\tt barriers-0.9} to construct the barrier tree from an energy
sorted list of spin configurations in linear time. The algorithm
explicitly constructs the basins $B(s)$ and subtrees $\mathfrak{T}(s)$
\cite{Flamm}.

\begin{figure}[t]
  \centerline{\epsfig{file=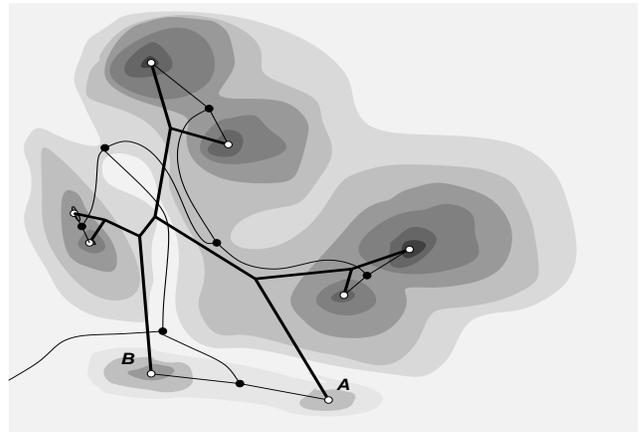,width=0.97\linewidth}} \par
  \caption{Unrooted phylogenetic tree of extant species (white dots)
  obtained from
  the minimum fitness paths, shown by thin lines with saddle points
  (ancestors) indicated by black dots, does not necessarily coincide with
  the tree obtained from clustering methods that are based on sequence
  similarity, shown here with thick lines. The gray intensity is
  proportional to the fitness value.}
  \label{fig:2Dland}
  \vspace{-0.4truecm}
\end{figure}

%%%  The connection to biology

The barrier tree can be viewed as a phylogenetic tree with a single common
ancestor at the root. The evolutionary process leading to the extant
species (i.e., the tips of the tree) is an adaptive walk on the
rugged fitness landscape (\ref{H_p}).  Interestingly, according to
definition (\ref{eq:saddle}) a subtree connecting two tips corresponds to
the evolutionary path of {\em minimum fitness cost} which could be regarded
as a generalization of the maximum parsimony principle \cite{TreeBooks} to
rugged fitness landscapes. 
This differs from the usual approach in
molecular biology, where a flat fitness landscape and a diffusive behavior
in sequence space is assumed to justify the reconstruction of phylogenetic
trees based solely on sequence similarity, i.e., configurational
overlap. 
In Fig.~\ref{fig:2Dland} we give an example in which the
distance-based tree and the barrier tree have different topologies. In
particular, the species labeled $A$ and $B$ are much closer in the
barrier tree than in the maximum parsimony tree.  It is interesting to note
that barrier trees can be defined in a meaningful way also for continuous
energy surfaces. Their nodes are the local minima and the saddle points
satisfying Eq.~(\ref{eq:saddle}), while the edges can be associated with
saddle connections along which the energy varies monotonically.

%%% Computational details

The definition of barrier trees becomes more complicated if the energy
function is degenerate as in the case of $p$-spin models with even $p$,
where $\mathcal{H}_p(s)=\mathcal{H}_p(-s)$.  The appropriate definition of
the barrier tree is obtained by identifying the saddle points $s$ and $s'$
with the same interior node of the tree provided (i) they have the same
energy and (ii) they are connected by a path along which the energy does
not exceed $\mathcal{H}_p(s)=\mathcal{H}_p(s')$.  In the non-degenerate
case the trees are almost always binary.  Degeneracies can occur also for
geometric reasons since the same saddle point can connect more than two
basins. One may, however, ``expand'' a non-binary interior node into a
sequence of binary nodes at the same height. We use this technical trick
here to simplify the computations. This procedure is justified because it
affects only the saddle points above the saddle $\tilde s$ that connects
the ground-state and its mirror image, i.e., it affects only the nodes with
large basins, far beyond the regime where the tree exhibits self-similarity.

One important aggregate characteristic of a tree is the size-frequency
distribution of its saddles or subtrees $\psi(w)$, where the size $w=w(s)$ is,
in the simplest case, just the number of local minima or tips in $B(s)$.  
It is instructive to consider first a few examples of simple ideal trees for
which $\psi(w)$ can be calculated analytically. E.g., a symmetric
binary tree of depth $m \geq 1$ has $2^m-1$ nodes, of which $2^{m-1}$ are
tips and the remaining $2^{m-1} - 1$ are saddles. It can be easily shown
that there are $2^{m-k-1}$ saddles with sizes $w=2^k; k=1,\ldots,m-1$, so
that
\begin{equation}
  \psi(w) = \left\{
    \matrix {\frac{1}{w} \left ( 1 - 2^{1-m} \right )
                    &\textrm{if } \log_2 w \in\mathbb{N} \cr
      0             &\textrm{if } \log_2 w \notin\mathbb{N} .  \cr
    }
  \right.
\end{equation}
The other extreme is the asymmetric binary tree in which every
left child is a tip. There are $2m-1$ nodes: $m$ tips and 
one saddle with size $w$, $1\le w\le m-1$. Hence
\begin{equation}
  \psi(w) = \frac{1}{m-1} \qquad \textrm{ for } 2\le w\le m .
\end{equation}

\begin{figure}[t]
  \centerline{\epsfig{file=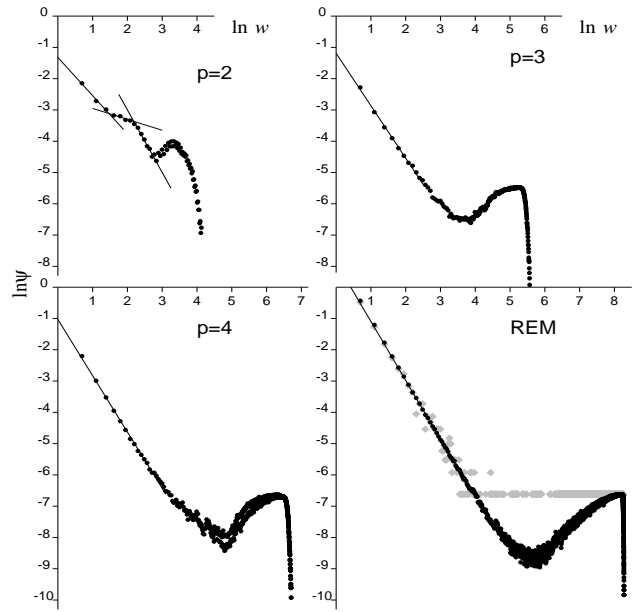,width=0.97\linewidth}}\par
  \caption{Log-log plots of the average frequency of subtrees (saddle
  points) with different numbers of tips (minima) for the $p$-spin
  landscapes with $N=18$, and the REM with $N=16$, averaged over 300
  landscapes each.  For the REM, a size-frequency distribution for a single
  landscape is shown as gray squares to demonstrate the scatter.}
\label{fig:saddle}
\vspace{-0.4truecm}
\end{figure}

We turn now to the analysis of the complex trees associated to the random
energy function (\ref{H_p}) as produced by the program {\tt
barriers-0.9}. In this case the average number of tips increases
exponentially with the number of spins $\mbox{e}^{\alpha_p N}$, where
$\alpha_p$ increases from $\alpha_2 = 0.199$ to $\alpha_\infty=
\ln 2$ \cite{MPV}. Log-log plots of the average size-frequency
distributions of subtrees for particular $p$-spin models and the REM are
shown in Fig.~\ref{fig:saddle}. In all cases the data are very well fitted
by straight lines with negative slopes in the regime of high frequencies,
suggesting then a power law form
\begin{equation}\label{power}
 \langle \psi(w) \rangle \sim w^{-D}
\end{equation}
where $D$ can be viewed as the fractal dimension of the barrier tree and
hence of the phase space of the disordered system.  This result points out
a very large number of subtrees with a few tips and a very small number of
subtrees with many tips. Moreover, it implies that there is no
characteristic number of tips within subtrees.  
As illustrated by the gray data points in the
panel for the REM, the same scaling law seems to hold true for the
size-frequency distribution of a single instance of the landscape as well.
The asymmetric scattering of points observed in the low frequency regime,
i.e., $\langle\psi(w)\rangle\ll1$, is due to the few high energy sadddle
points near the root of the tree.  It is interesting to note that many
frequency distributions of taxonomic units containing various numbers of
subunits (e.g., species per genus) were found to be well described by power
laws \cite{BioTrees}.  It should be stressed that the scaling law
(\ref{power}) is by no means a mere consequence of the existence of an
underlying tree structure, as it is clear from the asymmetric binary tree
example discussed before as well as from the study of discrete branching
processes which generate different forms of size-frequency distributions
\cite{DynamicTrees}.

The average size-frequency distribution for the SK model displays a rather
distinct behavior pattern which seems to indicate the existence of two
different types of self-similar structures at different levels of the tree.
These structures are characterized by straight lines with distinct slopes,
being joined by a short, almost flat curve corresponding to a crossover
regime between the dominant structures. The first slope appears to be
around $D_1\approx1.4$, the second slope is $D_2\approx2$. These results,
however, must be taken with caution since the size-frequency statistics is
greatly impaired by the fact that the trees are typically very small in
this model.  For instance, for $N=18$ the average number of tips is about
$50$ in the SK model as compared to the $\sim 10^5$ tips in the REM.

\begin{figure}[t]
  \centerline{\epsfig{file=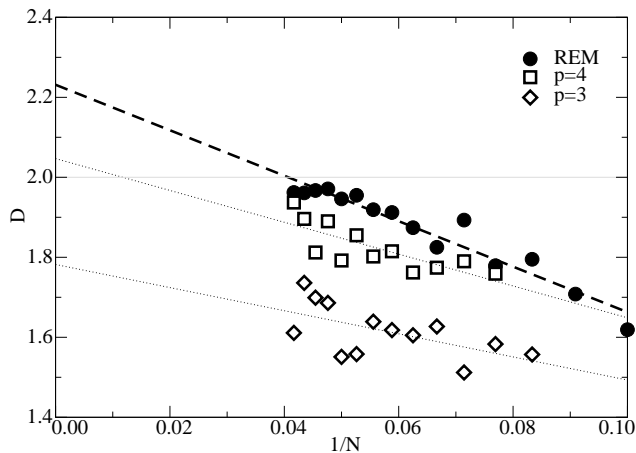,width=0.97\linewidth,clip=}}
  \par
  \caption{Dependence of the exponents $D$ on the reciprocal of
the  number of spins $N$. The data for $p=2$ do not allow for the unambiguous 
estimate of an exponent $D$.}
\label{fig:DxN}
\vspace{-0.4truecm}
\end{figure}

%%% Conclusions

In addition to being a {\it global} measure for the characterization of
rugged landscapes, to be contrasted with local measures such as the
correlation length \cite{Corrlen,Stadler:96b}, the parameter $D$ also
yields a measure of diversity since $D$ is higher in systems where subtrees
with one or a few tips are more numerous.  An attempt to estimate the
fractal dimension $D$ for infinite system sizes is presented in
Fig.~\ref{fig:DxN}. From these data one cannot discard the possibility that
for very large systems the exponents will converge to $D=2$ independently
of $p$, in which case the value of $D$ might be seen as an index that
characterizes the universality class of the $p$-spin landscapes.  Since the
exact construction of barrier trees is at present feasible for systems of
sizes up to $N = 24$ only, the estimate of $D$ for larger systems must
resort to a stochastic approach, probably in the spirit of the coalescent
theory of population genetics \cite{TreeBooks}, which focuses on the
genealogy of a few sampled individuals rather than on the family tree of
the entire population.

The replica theory predicts a similar hierarchical structure for the space
of pure states, consisting of clusters within clusters. In particular, the
probability density that a cluster has weight $W$ (roughly the fraction of
pure states within it) is \cite{MPV}
\begin{equation}\label{replica}
f(W) = \frac{W^{y-2} \left ( 1 - W \right )^{-y}}{\Gamma (y) \Gamma ( 1 -y )}
\end{equation}
which for small weights reduces to a power law $f(W) \sim W^{y-2}$. Here $y
\in (0,1)$ is a complicated function of the physical parameters $p$ and the
temperature $T$.  For instance, for the REM one has $y = 1 - T/T_c$ where
$T_c = \left ( 4 \ln 2 \right)^{-1/2}$.  As the replica pure states space
is a rather abstract construct (e.g., only a few low-energy local minima
are pure states and the definition of clusters does not involve the notion
of saddle points) a direct comparison with our results is not evident;
albeit it is highly desirable since we are not aware of any attempt to
verify Eq.~(\ref{replica}) via numerical simulations.

%%%%%%%%%%%%%%%%%%%%%%%%%%%%%%%%%%%%%%%%%%%%%%%%%%%%%%%%%%%%%%%%%%%%
{\small {\bf Acknowledgements.} The work of J.F.F. is supported in part by
CNPq and FAPESP, Proj.\ No.\ 99/09644-9. This research was performed during
a stay at ZIF in Bielefeld in May 2001 as part of the working group {\em
The Sciences of Complexity: From Mathematics to Technology to a Sustainable
World}.}

% References

\end{document}